\documentclass[final,5p,tightenlines,times,twocolumn,sort&compress]{elsarticle}

\usepackage{amsmath}

\DeclareMathSymbol{\mathbbH}{\mathord}{AMSb}{"48}
\usepackage{amssymb}
\usepackage{bbold}
\usepackage{bm}
\usepackage{color,graphicx}
\usepackage{slashed}
\usepackage{comment}
\usepackage{hyperref}
\usepackage{subcaption}
\usepackage{makecell}
\usepackage[dvipsnames]{xcolor}
\usepackage{pifont}
\usepackage{caption}
\usepackage{float}
\usepackage{array}
    \newcolumntype{P}[1]{>{\centering\arraybackslash}p{#1}}
    \newcolumntype{M}[1]{>{\centering\arraybackslash}m{#1}}

\DeclareMathSymbol{\mathbbE}{\mathord}{AMSb}{"45}

\journal{Physics Letters B}

\begin{document}

\renewcommand\labelitemi{$\vcenter{\hbox{\tiny$\bullet$}}$}
\begin{frontmatter}

\title{Decoding the proton's gluonic density with lattice QCD-informed machine learning}

\affiliation[anl]{%
   organization={High Energy Physics Division, Argonne National Laboratory},
   city={Lemont},
   postcode={IL 60439},
   country={USA}
}

\affiliation[msuphys]{%
   organization={Department of Physics and Astronomy, Michigan State University},
   city={East Lansing},
   postcode={MI 48824},
   country={USA}
}

\affiliation[msucmse]{%
    organization={Department of Computational Mathematics, Science and Engineering, Michigan State University}, 
    city={East Lansing}, 
    postcode={MI 48824},
    country={USA}
}

\author[anl]{Brandon Kriesten}
\ead{bkriesten@anl.gov}

\author[msuphys,msucmse]{Alex NieMiera}
\ead{niemiera@msu.edu}

\author[msuphys,msucmse]{William Good}

\author[anl]{T.~J.~Hobbs}

\author[msuphys]{Huey-Wen Lin}

\begin{abstract}
We present a first machine learning-based decoding of the gluonic structure of the proton from lattice QCD using a variational autoencoder inverse mapper (VAIM). Harnessing the power of generative AI, we predict the parton distribution function (PDF) of the gluon given information on the reduced pseudo-Ioffe-time distributions (RpITDs) as calculated from an ensemble with lattice spacing $a\! \approx\! 0.09$ fm and a pion mass of $M_\pi\! \approx\! 310$ MeV. 
The resulting gluon PDF is consistent with phenomenological global fits
within uncertainties, particularly in the intermediate-to-high-$x$ region where lattice data are most constraining. 
A subsequent correlation analysis confirms that the VAIM learns a meaningful latent representation, highlighting the potential of generative AI to bridge lattice QCD and phenomenological extractions within a unified analysis framework.
\end{abstract}
\end{frontmatter}

%
%
\noindent ANL-197798, MSUHEP-25-013, INT-PUB-25-020

\section{Introduction}
\label{sec:intro}

\begin{figure}[!h]
    \centering
    \includegraphics[width=0.925\linewidth]{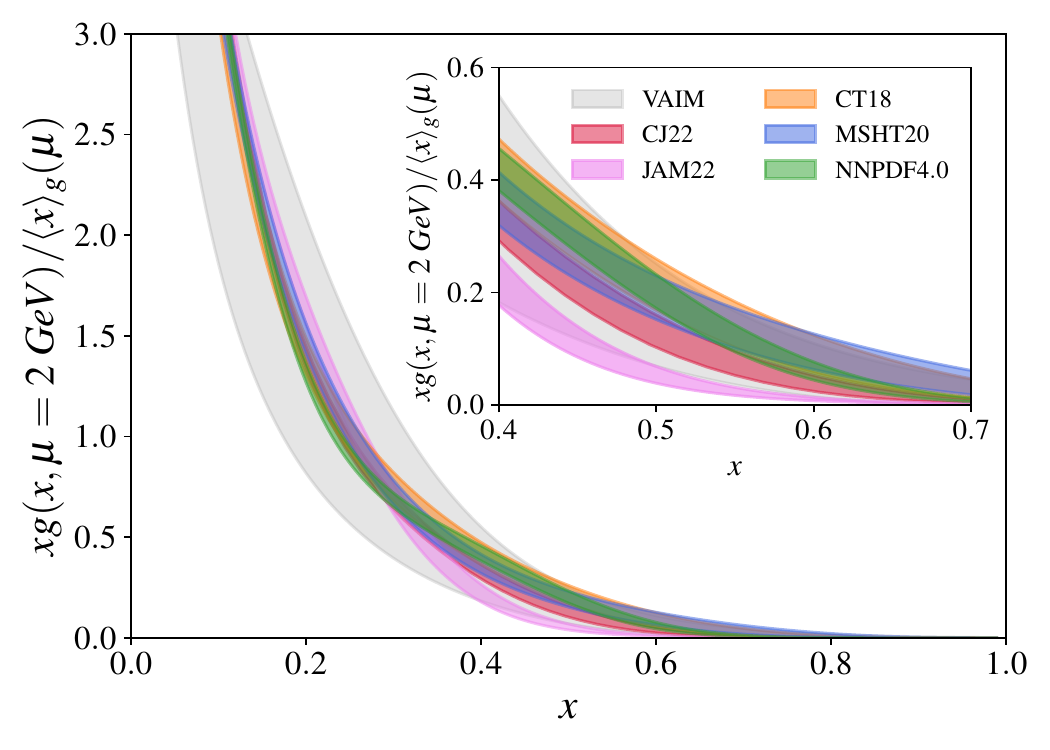}
    \includegraphics[width=0.925\linewidth]{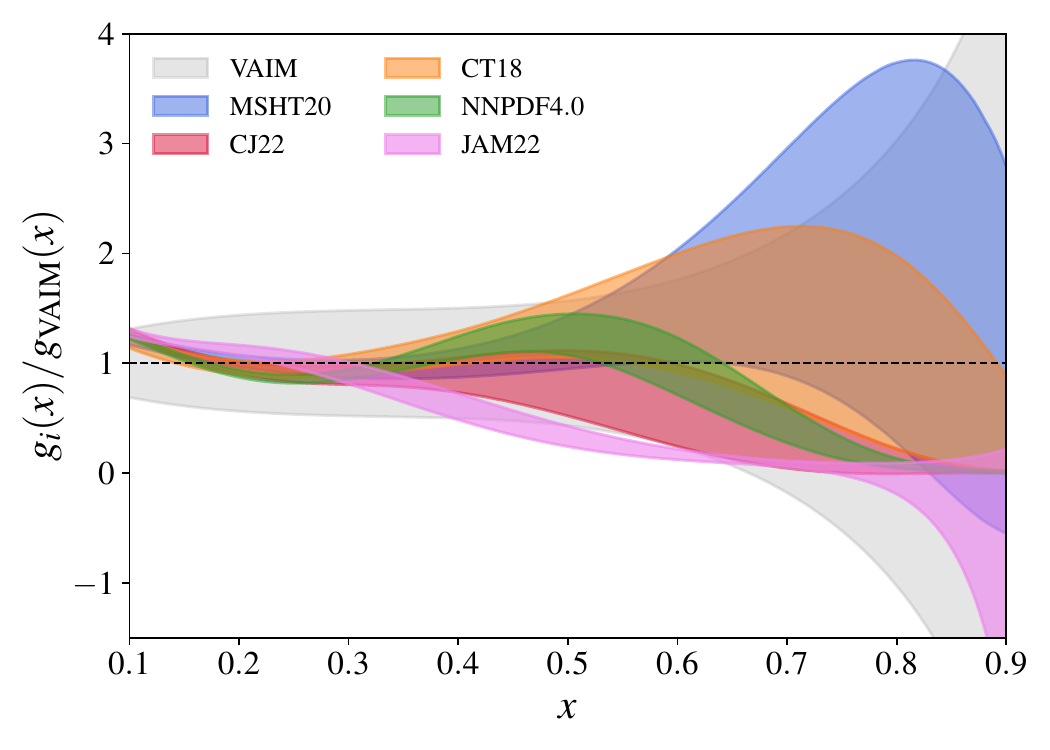}
    \caption{
    Upper: Comparison of the proton’s gluon PDF, $xg(x,\mu)/\langle x \rangle_g$, at $\mu = 2\, \mathrm{GeV}$ as predicted by a trained VAIM (gray shaded band), against global QCD fits including CJ22~\cite{Cerutti:2025yji} (red band) with multiplicative higher-twist corrections, CT18~\cite{Hou:2021} (orange band), JAM22~\cite{Cocuzza:2022jye} (pink band), MSHT20~\cite{Bailey:2020ooq} (blue band), and NNPDF4.0~\cite{NNPDF:2021njg} (green band). 
    Lower: Relative uncertainties for each gluon PDF as the ratio of the error band to the VAIM central value.}
    \label{fig:predicted-PDF}
\end{figure}

The constraining power of high-energy experiments is closely connected to knowledge of the partonic substructure of the colliding protons, including their gluonic density. At hadron colliders, gluon-initiated processes drive flagship channels such as Higgs boson production via gluon fusion ($g g \rightarrow H$), top-antitop pair creation ($gg\rightarrow t\bar{t}$), and the multi-TeV dijet spectra ($gg \rightarrow jj$) that anchor many new-physics searches~\cite{Ethier:2021bye, Gao:2022srd, Costantini:2024xae, Amoroso:2022eow}. Similarly, information on the gluon is central to efforts to better understand hadronic structure in its own right, from exclusive quarkonium photoproduction ($\gamma p \rightarrow J/\psi p$) in an upgraded 22-GeV CEBAF at JLab~\cite{Accardi:2023chb}, to small-$x$ observables that will be mapped with unprecedented precision at the forthcoming EIC~\cite{Accardi:2012qut}.

The gluonic structure of the proton is encoded by its corresponding gluon quantum correlation functions \cite{Kriesten:2019jep, Kriesten:2021sqc}. We concentrate in this study on the collinear unpolarized gluon parton distribution function (PDF), which depends upon a partonic momentum fraction, $x$, and resolution scale, $Q^{2}$.
To constrain the PDFs over as much of their flavor and $x$ dependence as possible, global fits exploit large sets of hadronic data covering a substantial kinematic range. This approach can be limited by regions of comparative data sparsity, or potential incompatibilities among experimental measurements. In parallel, lattice QCD calculations have matured to the point of providing complementary, \textit{ab initio} information on hadron structure in regimes for which experimental data may be more limited. As a result, recent global analyses have explored scenarios for including lattice QCD information alongside experimental data~\cite{Lin:2017stx,Hou:2022onq, Ablat:2025xzm}.

Advances in the theoretical framework of Large Momentum Effective Theory (LaMET)~\cite{Ji:2013dva}, the pseudo-PDF method~\cite{Balitsky:2021bds} and other approaches have enabled direct access to the $x$-dependence of PDFs from Euclidean-space correlation functions. 
While substantial progress has been made in extracting quark PDFs using these techniques, lattice determinations of gluon distributions remain comparatively limited due to the poorer signal-to-noise ratios associated with gluonic matrix elements, introducing considerable statistical challenges.
Nonetheless, recent studies have successfully applied these approaches to probe gluon distributions in the kaon~\cite{Salas-Chavira:2021wui, NieMiera:2025inn}, the pion ~\cite{Fan:2021bcr, Good:2023ecp}, and the nucleon ~\cite{HadStruc:2021wmh, Fan:2022kcb, Good:2024iur, Good:2025daz, HadStruc:2022yaw}. 
The majority of these studies have used the pseudo-PDF~\cite{Balitsky:2021bds} method 
and fit lattice data to phenomenologically inspired forms such as $f(x) = x^\alpha(1-x)^\beta$.

Unraveling the PDFs from hadronic data or lattice QCD information ultimately amounts to the solution of a complex inverse problem, wherein a wide range of underlying PDF behaviors may be statistically compatible with constraints imposed by the lattice and/or phenomenology.
Recent advances in exploring this inverse problem include the use of neural networks reconstructions~\cite{Karpie:2019eiq, DelDebbio:2020rgv, Dutrieux:2024rem, Khan:2022vot, Chowdhury:2024ymm, Zhang:2020gaj, Boyda:2022nmh}.
This work builds on these efforts by providing the first decoding of the proton’s gluon distribution from lattice QCD using a machine learning (ML)-based variational autoencoder inverse mapper (VAIM)~\cite{Almaeen:2021, Grigsby:2020auv, Kriesten:2023uoi, Almaeen:2024guo}, opening a new route for determining hadron structure through generative AI-powered unified pheno-lattice global analysis~\cite{Kriesten:2023uoi,Almaeen:2024guo}. Below, we exhibit our decoded gluon PDF in Sec.~\ref{sec:PDFresults} and present the details of our chosen lattice and ML methods in Sec.~\ref{sec:background}, before discussing the treatment of uncertainties and concluding in Sec.~\ref{sec:uq} and~\ref{sec:conc}, respectively.

\section{Decoded gluon PDFs}
\label{sec:PDFresults}

In Fig.\ref{fig:predicted-PDF}, we present the first prediction of the gluon PDF obtained from a trained VAIM, which learns a latent representation of reduced pseudo-Ioffe-time distributions (RpITDs). Post-training, this VAIM may be supplied with lattice-calculated values of the RpITDs, from which a gluon PDF may then be generated as discussed in Sec.~\ref{sec:inv_map}.
The resulting distribution is shown as a gray shaded band and is compared with five representative global analyses: CJ22~\cite{Cerutti:2025yji} (red band) with multiplicative higher-twist corrections, CT18~\cite{Hou:2021} (orange band), JAM22~\cite{Cocuzza:2022jye} (pink band), MSHT20~\cite{Bailey:2020ooq} (blue band), and NNPDF4.0~\cite{NNPDF:2021njg} (green band).
The upper panel displays the normalized gluon distribution, $xg(x,\mu\! =\! 2\,\mathrm{GeV})/\langle x \rangle_g(\mu)$, illustrating that all phenomenological fits comfortably agree with our VAIM prediction within their nominal uncertainties; we note a slight exception with JAM$22$, which largely agrees as well over all $x\! \ge 0.1$ outside the narrow range $0.5\! \lesssim\! x\! \lesssim\! 0.6$. 
We restrict Fig.~\ref{fig:predicted-PDF} to the high-$x$ region, as the most meaningful constraints occurs over $0.2 \lesssim x \lesssim 0.7$, where the lattice RpITDs are most sensitive to the underlying gluon distribution. 
This sensitivity arises from the behavior of the dimensionless Ioffe-time variable, $\nu = z \cdot P_z$, which informs the shape of the RpITD.
At moderate $\nu$ (corresponding to $0.2\! \lesssim\! x\! \lesssim\! 0.7$), the signal-to-noise ratio remains high and the RpITDs retain their greatest constraining power over the gluon PDF. 
Probing smaller $x$ demands large $\nu$, pushing the correlator to long Euclidean times where the signal decays exponentially and the matrix elements become susceptible to higher-twist contamination. 
At the opposite extreme, even larger $x$ corresponds to still smaller $\nu$; here, the RpITD signal goes to one, and the short-distance matrix elements become increasingly sensitive to discretization effects and lattice artifacts, limiting the attainable precision.

Thus, the $x$-region of $0.2\! \lesssim\! x\! \lesssim\! 0.7$ provides the most informative window for probing the lattice-informed gluon PDF. 
We highlight this region in the inset in Fig.~\ref{fig:predicted-PDF}, which reveals particularly strong agreement between the VAIM result and several global fits. 
Given the agreement between the VAIM prediction and the majority of the phenomenological fits, we have confidence in the reliability of our lattice-informed VAIM prediction. 

The lower panel of Fig.~\ref{fig:predicted-PDF} shows the uncertainty bands of several representative phenomenological PDFs, normalized to the central value of the VAIM prediction for easier comparison of the uncertainties. 
This collection of global fits, which includes CJ$22$, CT$18$, JAM$22$, MSHT$20$, and NNPDF$4.0$, exhibits uncertainties that are broadly consistent with or narrower than those of the VAIM result up to roughly $x\! \approx\! 0.5$, suggesting good statistical agreement, especially in the high-$x$ region most constrained by the lattice RpITD information.
Although direct comparisons among the uncertainties of the VAIM prediction and phenomenological PDFs should be made with care due to methodological differences among different uncertainty prescriptions, the overall level of agreement supports the statistical consistency of the lattice-informed VAIM prediction.
At small $x$, the VAIM shows a significant absolute uncertainty, reflecting the model’s flexible exploration of multiple allowed solutions where the lattice constraints are comparatively weak. 

\begin{figure}[htbp]
    \centering
    \includegraphics[width=0.95\linewidth]{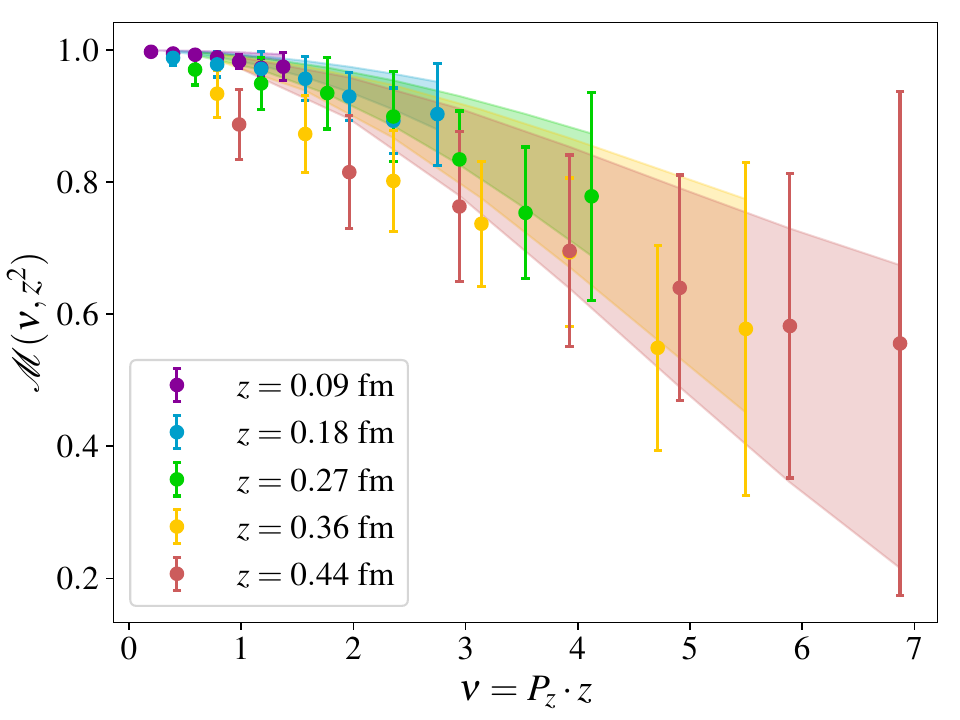}
    \caption{Closure test comparing lattice RpITDs (data points) to the RpITDs reconstructed from the VAIM-predicted PDF parameters (shaded bands), shown as a function of Ioffe-time. 
    Each color corresponds to a fixed spatial separation $z$, with the model tested across all combinations of $z \in [1a,5a]$ and $P_z \in \frac{2 \pi}{a L} \times [1, 7]$.}
    \label{fig:closure-test}
\end{figure}

In Fig.~\ref{fig:closure-test}, we present the results of a closure test comparing the input lattice RpITDs with those computed from the gluon PDFs predicted by the trained VAIM model. 
Each data point corresponds to the lattice RpITD at a specific combination of spatial separation, $z \in [1a, 5a]$, and momentum, $P_z \in \frac{2 \pi}{a L}  \times [1, 7]$, where $a$ is the lattice spacing and $L$ is the spatial extent of the lattice, as a function of the dimensionless Ioffe-time, $\nu$. 
The shaded bands represent the RpITDs reconstructed from the PDF parameters predicted by the trained VAIM, with different colors corresponding to fixed values of $z$. 
Overall, the reconstructed RpITDs exhibit good agreement with the lattice data, with an average point-by-point deviation below $1\sigma$ and a maximum deviation within $2\sigma$, the latter occurring primarily at small values of $\nu$, where discrepancies are more noticeable. 
In particular, we find that the lattice data corresponding to the lowest momentum, $P_z \approx 0.44$ GeV,
displays a rapid decay that is not well described by the expected matching relationship sketched below.
This behavior has also been reported in previous studies~\cite{Good:2024iur, NieMiera:2025inn}, though it can be neglected at the current level of statistics.

\section{Methodology and Formalism}
\label{sec:background}
In this section, we outline the lattice QCD framework used in this manuscript and review the methodology employed for the lattice-informed inverse mapping procedure. 

\subsection{Lattice RpITDs}
\label{sec:lattice}
In this work, we utilize the ground-state matrix elements obtained by the MSULat group in Ref.~\cite{Fan:2022kcb}.
We use the matrix elements computed on a lattice featuring $N_f = 2 + 1 + 1$ flavors of highly improved staggered quarks (HISQ)~\cite{Follana:2006rc}, generated by the MILC Collaboration~\cite{MILC:2012znn}, with a lattice spacing of approximately $a \approx 0.09$ fm and a pion mass of about $M_\pi \approx 310$ MeV.
We refer interested readers to Table $1$ of Ref.~\cite{Fan:2022kcb} for additional details regarding the lattice setup and the procedure used for the extraction of these matrix elements.

\noindent \textbf{Pseudo-PDF Method.}
Formally, the RpITD is defined as the double ratio of fitted ground-state matrix elements defined in Ref.~\cite{Fan:2022kcb},
\begin{equation} \label{eq:RpITD}
    \mathcal{M}(\nu, z^2) 
    = \frac{M (z \cdot P_z,z^2)/ M(0 \cdot P_z, 0)}{ M(z \cdot 0, z^2) / M(0 \cdot 0, 0)}\ ,
\end{equation}
where $\nu = P_z \cdot z$ is the dimensionless Ioffe-time and $M(z \cdot P_z, z^2)$ denotes the fitted ground-state matrix element corresponding to a Wilson line of length $z$ and hadron momentum, $P_z$. 
This double ratio is defined such that the RpITD is normalized to $1$ at $\nu = 0$, acting to cancel renormalization and kinematic factors and reducing some of the lattice systematics and ultraviolet divergences.
Crucially for training our VAIM, we access the gluon PDF using the pseudo-PDF method~\cite{Balitsky:2021bds}, which relates the RpITD above,
$\mathcal{M}(\nu, z^2)$, to the lightcone gluon PDF, $g(x, \mu^2)$, through a perturbative matching relation:
\begin{equation} \label{eq:matching-relationship}
    \mathcal{M}(\nu, z^2)
    = \int_0^1 dx \; \frac{x g(x, \mu^2)}{\langle x \rangle_g (\mu)} R_{gg}(x\nu, z^2\mu^2) + \mathcal{O}(z^2 m^2, z^2 \Lambda_{\text{QCD}})\ ;
\end{equation}
here, $\mu$ is the renormalization scale in the $\overline{\text{MS}}$ scheme, $\langle x \rangle_g (\mu) = \int_0^1 dx \; x g(x, \mu^2)$ is the gluon momentum fraction, and $R_{gg} (x\nu,z^2\mu^2)$ is the gluon-gluon matching kernel defined in Ref.~\cite{Balitsky:2021bds}.  
The residual power-suppressed corrections, of order $\mathcal{O}(z^2 m^2, z^2 \Lambda_{\text{QCD}}^2)$, arise from the finite spatial separation of the Wilson line in the Euclidean correlation function; here, $m$ is the nucleon mass reflecting finite-mass effects, and $\Lambda_{\text{QCD}}$ is the QCD confinement scale characterizing nonperturbative strong-interaction dynamics. 
These correction vanish as the separation approaches the light-like limit, $z^2 \rightarrow 0$. 
In this study, we neglect the quark-gluon mixing kernel present in the definition of Eq.~\ref{eq:matching-relationship} in Ref.~\cite{Balitsky:2021bds}, an approximation supported by previous lattice QCD calculations of gluon PDFs~\cite{Fan:2022kcb, Delmar:2023agv, Fan:2021bcr}, wherein the systematic impact of quark contributions was found to be significantly smaller than current statistical uncertainties.

\subsection{Inverse Mapping Methodology}
\label{sec:inv_map}
We employ the variational autoencoder inverse mapper (VAIM)~\cite{Almaeen:2021, Grigsby:2020auv, Kriesten:2023uoi, Almaeen:2024guo} to map lattice QCD observables, such as the RpITDs as described above, into their underlying partonic densities. The VAIM is built on the foundations of a variational autoencoder (VAE) \cite{Kingma:2013}. 
A conventional VAE learns to generatively map a latent prior, $\zeta$, to the input feature space, $\xi$, during training. 
The VAIM in our study augments this architecture with a secondary pathway of information: the measured observable, $\mathcal{O}$. During training, $\mathcal{O}$ is injected into the latent space and concatenated with $\zeta$. The joint latent vector, $(\mathcal{O},\zeta)$, is then decoded back to a reconstruction of the input feature space, $\xi$. 
By partitioning the latent manifold into an observable space and a generative subspace, a VAIM thus enables the inverse mapping from $\mathcal{O} \rightarrow \xi$, where the observable provides the forward mapping partial constraints, while the complementary degrees of freedom are captured in $\zeta$, turning the ill-posed inverse mapping into a well-posed, end-to-end problem. 

\noindent \textbf{Training Data.}
In training our VAIM to connect $x$-dependent behaviors of the gluon PDF to the RpITDs of Sec.~\ref{sec:lattice}, we adopt an explicit parameterization; namely, we take
\begin{equation} \label{eq:pdf-parameterization}
    \begin{split}
    \frac{x g(x, \mu^2)}{\langle x \rangle_g (\mu)} 
    = \frac{x^\alpha (1-x)^\beta (1 + \gamma \sqrt{x} + \delta x)}{\int_0^1 dx \; x^\alpha (1-x)^\beta (1 + \gamma \sqrt{x} + \delta x)}\ ,
    \end{split}
\end{equation}
where the denominator ensures normalization to the gluon momentum fraction, $\langle x \rangle_g(\mu)$.
This parameterization has been employed in previous VAIM studies~\cite{Kriesten:2023uoi}, and is comparable to functional forms used in phenomenological fits.
The training data for the VAIM are ({\it i}) the input feature space ($\xi$, in the discussion above) consisting of PDF shape parameters, $\alpha, \beta, \gamma, \delta$, and ({\it ii}) the observable space ($\mathcal{O}$), formed from RpITDs calculated according to Eq.~(\ref{eq:matching-relationship}). 
The PDF parameters are sampled from uniform distributions defined as: $\alpha \in \mathcal{U}(-0.9,1)$, $\beta \in \mathcal{U}(0,10)$,  and $\gamma, \delta \in \mathcal{U}(-5,5)$.
The 35 RpITDs are calculated using the input PDF parameters with the prescription as described in Sec.~\ref{sec:lattice}. We use a total of 400,000 data samples split into training/validation/test sets with ratios 80/10/10. 
The training PDFs are chosen such that they obey physics constraints such as positivity. 
The input parameters are also restricted such that the calculated RpITDs fall within the range $\mathcal{M}(\nu, z^2) \in [-0.5, 1]$. 
The upper bound enforces the normalization condition outlined in Sec.~\ref{sec:lattice}, where $\mathcal{M}(\nu=0, z^2) = 1$ by definition, while the lower bound reflects behavior observed in lattice QCD calculations, where RpITDs exhibit a monotonic, approximately exponential decay from unity at $\nu = 0$ toward zero at large Ioffe-time. 
Though small oscillations around zero may occur for large values of $\nu$, the RpITD is not observed to grow beyond its initial value.
Thus, these bounds help maintain physical consistency,  while loosely guiding the model toward the qualitative features observed in lattice data.

\noindent \textbf{Architecture Details.} 
The VAIM is composed of a pair of symmetric networks to ensure the balance of strength between the encoding to the latent manifold and decoding to the reconstructed input feature space. 
Each network contains four identical residual blocks (ResModules) (see description in \cite{Kriesten:2024are}) derived from ResNet-like architectures \cite{DBLP:journals/corr/HeZRS15}, which maintain stable gradient flow during backpropagation through skip connections. 
Each ResModule follows the pattern: input layer $\rightarrow$ Dense(units=1000, `dense1') $\rightarrow$ ELU $\rightarrow$ Dense(units=1000, `dense2') $\rightarrow$ Sum(input,`dense2') $\rightarrow$ ELU; where Dense($\cdot$) is a fully connected Dense layer, ELU is the exponential linear unit activation function, and Sum is a summation layer. 
The outputs of the decoder has 3 parallel heads to predict PDF parameters: (1) a Dense layer with a $\tanh$ activation to predict $\alpha$, (2) a Dense layer with a Softplus activation to predict $\beta$, and (3) a custom layer with Softplus activation to constrain $\gamma$ and $\delta$ to obey a basic positivity constraint $1 + \gamma + \delta > 0$. The encoder model predicts the 256-dimensional mean and standard deviations for the generative latent variable as well as a 35-dimensional observable vector.

The VAIM is trained using the Adam optimizer with a learning rate of $10^{-6}$, and early stopping to prevent overfitting. 
Alternative configurations ranging from two to five ResModules and 100 to 1000 units per layer were explored systematically; the four-block, 1000-unit architecture adopted here yielded the optimal $\chi^{2}$ statistic from a closure test on lattice QCD data.

\noindent \textbf{Generating Predictions.} 
After training, the VAIM can generate PDFs based on RpITD observables calculated on the lattice. 
Random points are sampled from the latent distribution, $\zeta$, and the lattice RpITDs, $\mathcal{M}(\nu,z^2)$, are fed into the observable space $\mathcal{O}$; this generates predicted PDF parameters which are then ensembled using the methods described in Sec.~\ref{sec:uq}.

%
\section{Treatment of uncertainties and correlations}
\label{sec:uq}
In this section, we describe the treatment of uncertainties within our VAIM framework and discuss Pearson correlations to inspect various pulls from lattice-calculated RpITDs on the PDF parameters and subsequent $x$-dependent shapes. 
\subsection{Uncertainty quantification}
\label{sec:treat}
Uncertainty quantification in machine learning and its relationship with physics uncertainties is a nascent subject \cite{Almaeen:2022imx, Kotz:2023pbu, Kriesten:2024ist, Kotz:2025lio}. In this work, we quantify uncertainties by a standard Monte Carlo (MC) ensembling method. 

The lattice RpITDs enter the analysis in the form of $N$ Jackknife samples. To interface these lattice data with the VAIM to generate predicted PDF parameters, we convert the Jackknife ensemble into a Gaussian MC ensemble using the following prescription: 
\begin{equation}
    \mathcal{O}_{\text{MC}} \sim \mathcal{N}\left(\overline{\mathcal{O}}_{\text{JK}} ; \sigma_{\text{JK}}^{2}(N-1)\right) ,
\end{equation}
where $\overline{\mathcal{O}}_{\text{JK}}$ is the Jackknife mean which is the same as the MC mean, and $\sigma^{2}_{\text{JK}}$ is the Jackknife variance which is scaled by a factor of $N-1$ to reproduce the $1\sigma$ confidence interval. 
The training of the VAIM is performed using MC replicas of theory calculated RpITDs; therefore, it is necessary that the lattice calculated inputs are converted to an MC ensemble, representative of the training distribution.

For each MC replica, we sample the generative latent vector from the variational posterior of the encoder model. The MC sample and the variational latent vector are concatenated and passed through the trained decoder model which returns a set of predicted PDF parameters: $(\mathcal{O}_{i},\zeta_{i}) \rightarrow \{\alpha_{i},\beta_{i},\gamma_{i},\delta_{i} \}$. Repeating this procedure yields an ensemble of parameter sets from which standard statistical measures can be calculated. With this methodology, we encapsulate the uncertainty inherent in the trained model and exploit the full degrees of variation in the architecture and input data. 

Several systematic uncertainties are not yet explored in this analysis: functional parameterization choice, the conversion of lattice Jackknife ensembles to MC ensembles, or the separation of \textit{aleatoric} and \textit{epistemic} uncertainties. 
Addressing these questions requires a dedicated study within an analysis framework assimilating lattice, ML and physics uncertainties simultaneously. 
Such a study is well-motivated but beyond the scope of the current manuscript.

\subsection{Correlations}
\label{sec:corr}
Beyond the MC ensemble calculation, we also use the Pearson correlation to inspect $x$-dependent regions of the PDFs most influenced by the lattice RpITD calculation. Novel treatments of explainability within ML networks such as those discussed in Ref.~\cite{Kriesten:2024are} are left to future investigation.

The analog of the Pearson correlation for a PDF-dependent quantity --- including the gluon PDF, $g_{i}(x)$, itself --- can be evaluated over the ensemble of MC replicas as
\begin{equation}
\label{eq:corr-def}
    \text{Corr}(X,Y) = \frac{\langle X Y \rangle - \langle X \rangle \langle Y \rangle}{\Delta X \Delta Y}\ ,
\end{equation}
where $\langle \cdot \rangle$ is the expected value of some PDF dependent quantity, and $\Delta \cdot$ is the associated uncertainty (normalization factor) given the standard master formulas defined in Refs.~\cite{Pumplin:2001ct, Pumplin:2002vw, Hou:2016sho}.

\begin{figure}[htbp]
    \centering
    \includegraphics[width=1.0\linewidth]{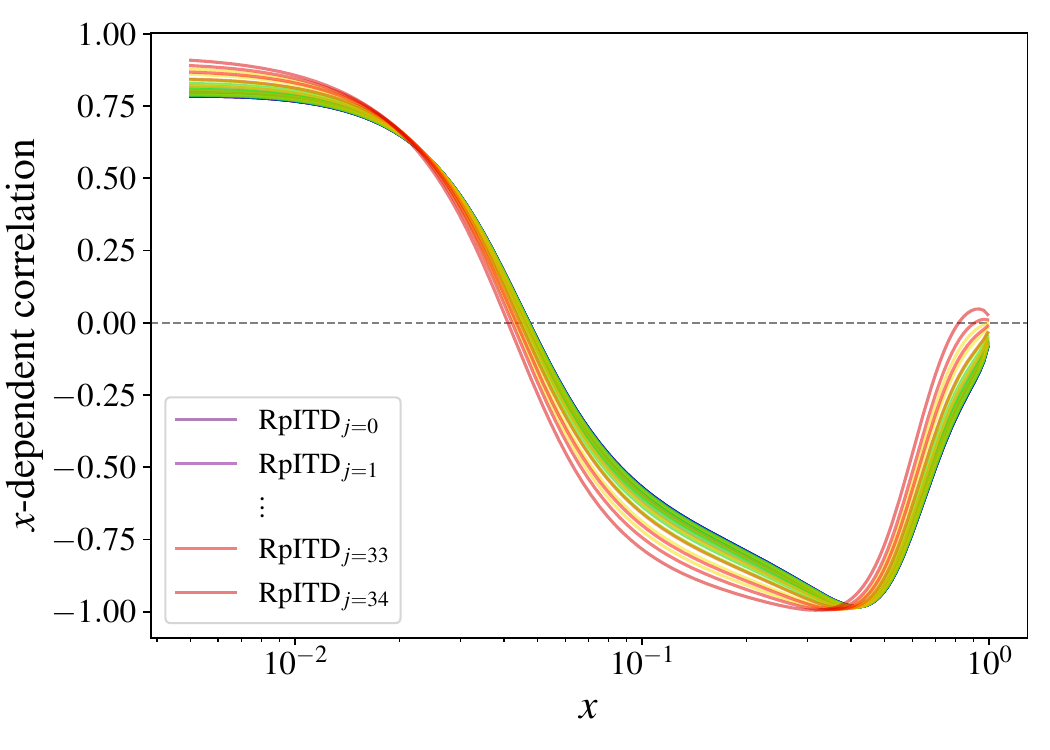}
    \caption{Pearson Monte Carlo correlations between the latent RpITDs and $x$-dependent gluon PDF as evaluated over an ensemble of trained model predictions for the normalized gluon PDF, $xg(x,\mu)/\langle x \rangle_g$.}
    \label{fig:correlations}
\end{figure}

In Fig.~\ref{fig:correlations}, we present the $x$-dependent Pearson correlation coefficients between each of the 35 predicted RpITDs and the normalized gluon distribution, $xg(x, \mu = 2 \text{ GeV})/\langle x \rangle_g(\mu)$. 
These correlations are computed across an ensemble of predictions generated by the trained VAIM. 
Each curve corresponds to a distinct RpITD, characterized by a unique combination of spatial separation, $z$, and momentum, $P_z$, and represents how variations in that specific matrix element influence the predicted gluon PDF at different values of the momentum fraction, $x$.
The color gradient encodes the index $j \in \left[0,34\right]$, which reflects increasing spatial separations, $z$, and momentum, $P_z$. 
The purple-to-blue lines ({\it e.g.}, $j=0,1,2$) correspond to the smallest values of $z$, while the yellow-to-red lines ({\it e.g.}, $j=32,33,34$) indicate the largest values. 
This color scheme is consistent with the $z$-dependent band coloring used in Fig.~\ref{fig:closure-test}, enabling a direct visual connection between regions of the RpITD input space and their contributions to the reconstructed gluon PDF.

Importantly, the bunching (fanning) of the curves at a given value of $x$ indicates the redundancy (complementarity) of the RpITDs in constraining the PDF. 
When the correlation lines are tightly bunched together around a common value, it suggests that many RpITDs share similar constraining power over the gluon PDF at that value of $x$. 
In contrast, the fanning out of the $x$-dependent correlations implies a diversity in the sensitivity of the RpITDs, with some values playing a more dominant role in that region of $x$.

%

%
\section{Conclusions}
\label{sec:conc}
In this study, we present a first extraction of the gluon PDF from lattice-calculated RpITDs with a variational autoencoder-based inverse mapper (VAIM) algorithm. We propagate uncertainties by exploiting the variational nature of VAIM and sampling the learned latent distribution. We achieve uncertainty bands that are comparable with phenomenological extractions for the large-$x$ region most connected to the RpITDs, and which widen at smaller $x$ beyond the reach of the lattice data. Within these uncertainties, the VAIM-extracted gluon PDF is consistent with phenomenological determinations from CJ22, CT18, JAM22, MSHT20, and NNPDF4.0. A Pearson correlation study between the latent RpITDs and the $x$-dependent gluon PDF confirms that the generative decoder of the VAIM indeed learns an accurate representation of the PDFs. 

Our results open a generative AI-based pathway for embedding lattice QCD observables directly into global analyses. Such analyses will be greatly aided by a comprehensive determination of epistemic uncertainties, encompassing both the parameterization dependence of the analytic fitting form as well as the lattice statistical and systematic uncertainties. Once the epistemic error budget is known, improvements in the statistical precision of the lattice calculations will translate directly into predictable reductions in the $x$-dependent PDF uncertainties. As such, we expect VAIM-based calculations like that presented above to serve as a rapidly-improvable basis to revisit the impact of forthcoming lattice information and assess its phenomenological implications.
%
%
%

\section{Acknowledgments}
The work of BK and TJH at Argonne National Laboratory was supported by the U.S. Department of Energy under contract DE-AC02-06CH11357.
The work of WG and AN is partially supported by U.S. Department of Energy, Office of Science, under grant DE-SC0024053 ``High Energy Physics Computing Traineeship for Lattice Gauge Theory''. 
The work of WG and HL is  partially supported
by the US National Science Foundation under grant PHY~2209424.
We gratefully acknowledge use of the Bebop supercomputer in the Laboratory Computing Resource Center at Argonne National Laboratory for the training of the VAIM models and related computations.
We thank the MILC Collaboration for sharing the lattices used to perform this study.
The LQCD calculations were performed using the Chroma software suite~\cite{Edwards:2004sx}.
This research used resources of the National Energy Research Scientific Computing Center, a DOE Office of Science User Facility supported by the Office of Science of the U.S. Department of Energy under Contract No. DE-AC02-05CH11231 through ERCAP;
facilities of the USQCD Collaboration, which are funded by the Office of Science of the U.S. Department of Energy,
and supported in part by Michigan State University through computational resources provided by the Institute for Cyber-Enabled Research (iCER).  
BK, AN and HL thank the Institute for Nuclear Theory at the University of Washington for its kind hospitality and stimulating research environment; this research was supported in part by the INT's U.S. Department of Energy grant No. DE-FG02- 00ER41132.
%
%
%

\bibliographystyle{elsarticle-num} 
\bibliography{refs}

\end{document}